\documentclass[camera,letterpaper,nomarginnotes,nonarrowgutter]{jpaper}
\usepackage[sort,compress]{cite}
\usepackage{amsmath,amssymb,amsfonts}
\usepackage{graphicx}
\usepackage{textcomp}
\usepackage{xcolor}
\usepackage{fancyhdr}
\usepackage{balance}
\usepackage{titling}
\usepackage[bookmarks=true,breaklinks=true,letterpaper=true,colorlinks,citecolor=blue,linkcolor=blue,urlcolor=blue]{hyperref}

\usepackage{listings}
\usepackage{xcolor}
\usepackage{lipsum}
\usepackage{cleveref}

\newcommand{\subtitle}[1]{\posttitle{\par\normalfont{#1}\par\end{center}}}

\title{\Large{AutoPRAC: Automating Attack Discovery for PRAC-Based Rowhammer Defenses using Model Checkers}}

\newcommand{\TODO}[1]{\textcolor{magenta}{TODO: #1}}

\newcommand{\NBO}{$\text{N}_\text{BO}$}
\newcommand{\Nmit}{$\text{N}_\text{Mit}$}
\newcommand{\ACTABO}{$\text{ACT}_\text{ABO}$}
\newcommand{\ACTDelay}{$\text{ACT}_\text{Delay }$}
\newcommand{\ACT}{$\text{ACT}$}
\newcommand{\ACTs}{$\text{ACTs}$}
\newcommand{\TRH}{$\text{T}_{\text{RH}}$}
\newcommand{\MaxACT}{$\text{Max}_{\text{ACT}}$}

 \author{Joyce Qu \quad Gururaj Saileshwar
 \vspace{-0.15in}
 \\\\\emph{University of Toronto}
 \vspace{-0.1in}}

\begin{document}
\maketitle


\thispagestyle{plain}
\pagestyle{plain}



\begin{abstract}
Per-Row Activation Counting (PRAC) in DDR5 is a specification to mitigate Rowhammer attacks by tracking activations per row and triggering mitigative refreshes when needed. However, the security of PRAC designs is currently evaluated using human-crafted attack patterns and we lack formal verification of their security properties, or automated techniques to detect implementation flaws. In this work, we present \textit{AutoPRAC}, the first automated technique to test the security of PRAC-based defenses using model checkers. AutoPRAC models PRAC implementations as bounded state machines and checks security-critical safety properties against a worst-case oracle attacker. If a property is violated, the framework produces a concrete counterexample trace corresponding to a successful attack.
Using AutoPRAC, we uncover a previously unreported flaw in MOAT, a state-of-the-art PRAC defense, in its counter-reset policy that allows up to 34 activations to go undetected above the Rowhammer threshold. Our results demonstrate that AutoPRAC can automatically discover subtle security flaws in Rowhammer mitigations and serves as an early-stage design aid for attack discovery on PRAC designs.
\end{abstract}

\section{Introduction}
JEDEC introduced Per Row Activation Counting (PRAC) in the DDR5 specification~\cite{jedec_ddr5_prac} to mitigate Rowhammer attacks~\cite{Rowhammer2014} using per-row activation counters and the Alert Back-Off (ABO) protocol. Subsequent works~\cite{Panopticon,QPRAC,MOAT,UPRAC,Chronus,MoPAC,Counterpoint,DemystifyingPRAC} proposed PRAC implementations that reduce its runtime overheads while preserving security. However, all of these defenses have so far been evaluated only against human-crafted attacks and without formal guarantees that stronger attacks do not exist. Moreover, these defenses may contain implementation bugs or overlooked design flaws that enable more effective attacks.


To address this gap, we propose AutoPRAC, a framework that uses bounded model checking (BMC), a formal verification technique, to test the correctness properties of PRAC designs. When a property is violated, AutoPRAC produces a concrete counterexample consisting of a sequence of row activations that forms a successful attack, allowing designers to analyze and fix the vulnerability. 
Thus, AutoPRAC serves as an automated attack discovery tool suitable for surfacing vulnerabilities at design-time.\footnote{
AutoPRAC uses bounded model checking, a formal verification technique, that allows it to explore the attack space only within the specified bounds.
While AutoPRAC is effective as a bug-finding tool, it cannot prove the security of a design beyond the analyzed bounds, in the absence of a discovered attack.}


Security vulnerabilities in PRAC designs typically arise in two ways: 
(1) a \emph{tracking failure}, where the counter values no longer accurately reflect the true disturbance to neighboring rows, or (2) a \emph{mitigation failure}, where mitigation is delayed long enough for activations to exceed the safe threshold. Across existing PRAC designs~\cite{Panopticon,QPRAC,MOAT,UPRAC,Chronus}, we identify two key design decisions that can lead to these failures: how rows are selected for mitigation, and how counters are reset. Prior works use a variety of mitigation-selection mechanisms, including FIFO queues~\cite{Panopticon}, priority queues~\cite{QPRAC}, counter-based structures~\cite{MOAT,Chronus}, or no queue at all~\cite{UPRAC}; some of these designs have been shown vulnerable to mitigation failures~\cite{QPRAC,MOAT}. Regarding counter resets, while most designs do so only when neighboring rows are mitigated, MOAT~\cite{MOAT} additionally resets counters during periodic refreshes; however, counter reset policies can introduce tracking failures if not carefully designed.

AutoPRAC models two complementary aspects of PRAC defenses corresponding to their primary failure modes: incorrect mitigation scheduling decisions and incorrect counter reset behavior. We formalize both as safety properties and develop verification models that check whether a PRAC implementation satisfies them under a worst-case attacker. We implement AutoPRAC using the C Bounded Model Checker (CBMC)~\cite{cbmc}.

However, bounded model checking has fundamental scalability challenges. Verifying security against standard Rowhammer attacks requires exploring activation sequences across an entire refresh window (tREFW) spanning almost 550,000 \ACTs{}. This leads to state space explosion, making exhaustive exploration computationally infeasible. We improve tractability by 
(1) initializing PRAC counters near the Alert-Back-Off threshold, reducing the number of simulated activations, and (2) representing counter states as a histogram instead of tracking individual rows, reducing the state space.

We evaluate AutoPRAC on four PRAC designs: UPRAC~\cite{UPRAC}, Panopticon~\cite{Panopticon}, QPRAC~\cite{QPRAC}, and MOAT~\cite{MOAT}. Within a solver time limit of 2 hours, AutoPRAC synthesizes worst-case attack patterns up to 2105 \ACTs{}. While solver timeouts prevent AutoPRAC from surpassing prior human-crafted attacks for most designs, it discovers a stronger attack than any previously known against MOAT. It exploits a flaw in MOAT's counter reset policy, allowing up to 34 activations beyond the \TRH.

\noindent
Overall, this paper has the following contributions: 
\begin{itemize} 
\item We propose AutoPRAC, the first model checking framework to formally evaluate the security of PRAC designs.
\item 
We propose two optimizations to reduce the length of attack simulation and the number of simulated rows, reducing the state space to improve tractability. 
\item
While AutoPRAC's scalability limitations prevent full verification of defenses, it serves as the first automated attack discovery tool for PRAC defenses. 
\item AutoPRAC produces stronger attacks than any previously known against MOAT, uncovering a flaw in its counter reset policy, highlighting its capability to find new vulnerabilities. 
\end{itemize}

\section{Background}

\subsection{Per Row Activation Counting (PRAC)}
JEDEC's DDR5 specification~\cite{jedec_ddr5_prac} introduces Per Row Activation Counting (PRAC) as an in-DRAM Rowhammer defense. Each DRAM row maintains an activation counter that increments on every activation. When a counter exceeds a back-off threshold (\NBO{}), the DRAM raises an Alert signal through the Alert Back-Off (ABO) protocol to request mitigation time from the memory controller. The controller then issues a fixed number (\Nmit{}) of Refresh Management (RFM) commands during which the DRAM performs mitigative refreshes.

The ABO protocol is non-blocking. After an Alert is raised, the specification permits up to 180ns before mitigative refreshes begin, allowing a few extra activations to be serviced (\ACTABO{}). After mitigation completes, the protocol enforces a delay of \ACTDelay{} activations before another Alert can be issued.
These additional activations between successive mitigations (\ACTABO{}+\ACTDelay{}) create opportunities for attacks~\cite{UPRAC}. An attacker can first prepare a set of rows whose counters are all at \NBO{}$-1$, then trigger successive ABO rounds, issuing activations in the gaps between mitigations. As a result, a target row can accumulate significantly more than \NBO{} activations before being refreshed within a tREFW.
Therefore, PRAC defenses must reason about the worst-case number of activations a row can receive beyond \NBO{} before mitigation (\MaxACT{}). To remain secure, the back-off threshold must satisfy:
\NBO{} + \MaxACT{} $<$ \TRH{}, 
where \TRH{} is the Rowhammer threshold, i.e., the minimum number of activations to a row required to induce a bit flip in the DRAM device.

\subsection{PRAC-based Rowhammer Defenses}
Several recent defenses~\cite{UPRAC,Panopticon,MOAT,QPRAC} propose implementations of the PRAC specification. These designs primarily differ in two aspects: (1) how rows are selected for mitigation, and (2) how activation counters are reset.


\textbf{Variation in mitigation policy.}
When multiple rows exceed the back-off threshold (\NBO{}), the PRAC specification does not define which row should be mitigated first, creating a design space for mitigation scheduling policies.
Panopticon~\cite{Panopticon} uses a FIFO queue to track rows pending mitigation: when a row crosses \NBO{}, it is inserted into the queue, and each RFM mitigates the row at the head. However, under PRAC's non-blocking ABO protocol, Panopticon is vulnerable to the Toggle+Forget and Fill+Escape attacks~\cite{QPRAC}, where queue overflows allow rows to escape tracking or mitigation.
UPRAC~\cite{UPRAC} avoids a queue and always mitigates the globally highest-activated row whenever any row crosses \NBO{}. While this avoids queue-based attacks, finding the maximum counter across the entire bank causes substantial overhead.
QPRAC~\cite{QPRAC} maintains a small priority queue ordered by activation counts. Rows with sufficiently high counts are inserted even when the queue is full, and RFMs mitigate the highest-priority row. This preserves accurate tracking even when the queue is full, preventing the attacks affecting Panopticon. Similarly, MOAT~\cite{MOAT} tracks 1 to 4 highly activated rows, same as the number of RFMs issued on Alerts, and mitigates them on RFMs; Chronus~\cite{Chronus} also maintains a table of highly activated rows beyond \NBO{}.

\textbf{Variation in counter reset policy.}
All PRAC defenses reset activation counters when a row is mitigated, since mitigative refresh restores the neighboring row's tolerance to Rowhammer. Some designs additionally reset counters during periodic refreshes (REF commands), which occur throughout a refresh window (tREFW). Since refresh restores charge in DRAM rows, resetting neighboring counters appears reasonable; however, incorrect reset policies can allow attackers to accumulate activations while evading tracking.
Among the designs we evaluate, only MOAT~\cite{MOAT} explicitly resets counters during periodic refreshes. UPRAC, Panopticon, and QPRAC implicitly assume counters are preserved across refreshes.

\subsection{Model Checking}
Model checking is a formal verification technique that systematically explores a system's reachable states to determine if a correctness property holds on all executions. Modern model checkers typically encode this problem as a SAT or SMT query. If a property is violated, the checker returns a counterexample, i.e., a concrete execution trace that demonstrates the violation.

\textbf{Bounded model checking (BMC)} BMC restricts verification to executions of length at most $k$. In general, BMC is useful as a bug-finding technique: it can detect violations within $k$ steps, but does not by itself prove correctness beyond that bound. BMC is well-suited to Rowhammer verification because attacks must complete within a bounded refresh window (tREFW). Thus, if $k$ is chosen to cover a full refresh window, verifying the absence of violations within $k$ steps can verify the security of a PRAC defense.
However, the main limitation of BMC is scalability. Tools such as CBMC~\cite{cbmc} unroll program execution up to bound $k$ and encode all possible executions into a SAT/SMT formula. As the number of states and nondeterministic choices increases, the search space grows rapidly, leading to the classical \textit{state explosion} problem. Although modern model checkers employ optimizations such as abstraction refinement, partial-order reduction, and incremental
solving, scalability remains a fundamental challenge.

\textbf{Model checking in hardware security.} 
Despite the limitations of BMC, prior work has successfully leveraged BMC for attack discovery in hardware security. Pensieve~\cite{Pensieve} used BMC to analyze speculative execution defenses and uncovered previously unknown vulnerabilities despite verifying executions of only up to 9 CPU cycles due to solver limits. This demonstrates the practical value of bounded model checking as a bug-finding tool for early-stage hardware designs and motivates its application to PRAC defenses.

\section{Verification Models for PRAC Defenses}
\textbf{Assumptions.} We develop verification models for PRAC-based mitigations assuming a worst-case attacker. We make the following assumptions in our model:
\begin{itemize}
    \item We consider an attack targeting an isolated bank. Activities in other banks may affect the targeted bank (e.g., an all-bank RFM causing a mitigation to a row that has not yet crossed \NBO{}), but these can be ignored to model a worst-case attack.
    \item We assume the attacker can activate arbitrary rows, observe mitigations, and issue activations as aggressively as possible throughout a tREFW. We also assume the attacker does not access the same row more than once consecutively, as that would result in a row buffer hit and not activate the row, which is not beneficial for the attacker.
    \item We do not model any pre-emptive mitigations to a row before its row activation counter reaches \NBO{}, as this is not defined in the PRAC specification and does not affect the worst-case analysis. This includes issuing  mitigations in the shadow of REF commands on tREFIs or on an RFM to another bank.
\end{itemize}

We model PRAC designs in C++ since prior academic PRAC implementations~\cite{MOAT,UPRAC,QPRAC,Chronus} are open-sourced as C/C++ models in simulators like Ramulator~2~\cite{ramulator2}, without corresponding RTL implementations. 
This level of abstraction is sufficient for our primary goal of identifying algorithmic design flaws. While RTL-level verification could uncover additional implementation bugs, we leave such verification to future work.

We separately model two security-critical design decisions in AutoPRAC: the mitigation policy (\Cref{sec:model_queue_design}) and the counter reset policy during refresh (\Cref{sec:model_counter_reset}). Separating them reduces the model checker's search space and makes it easier to attribute violations to a specific design choice. The modular structure also allows different mitigation policies and reset policies to be composed independently.

\subsection{Modeling Mitigation Policy (Queue Design)}
\label{sec:model_queue_design}

\Cref{lst:naive_loop} shows the core structure of our bounded model checking loop implemented in the C Bounded Model Checker (CBMC)~\cite{cbmc} with the CaDiCaL SAT solver~\cite{cadical}. The goal of the checker is to verify that no row exceeds the specified Rowhammer threshold of the device(\TRH{}) under a given PRAC design and parameter configuration.
Given a PRAC implementation, the model nondeterministically selects rows to activate, invokes the implementation's logic to increment counters and invokes its mitigation logic to determine whether a mitigative refresh should occur. The mitigation logic is abstracted behind an interface with two operations: tracking rows requiring mitigation and selecting which row to mitigate. This allows different PRAC implementations to be verified by swapping in different tracking and mitigation policies.

\begin{lstlisting}[caption={Naive PRAC model checking loop}, label={lst:naive_loop}, xleftmargin=1.5em,  xrightmargin=0.5em]
int counters[N] = {0}; // Initialize all counters
int t = 0; // time step
while (t < tREFW) {
    // Model checker chooses any valid row
    int row_to_activate = nondet_int(); 
    assume( is_valid(row_to_activate) );
    // Row activation
    counters[row_to_activate]++;
    mitigation.track(row_to_activate);
    t += tRC;
    if (mitigation.needed && mit_available(t)) {
        int row_to_mitigate = mitigation.mitigate();
        // Mitigation
        counters[row_to_mitigate] = 0; // Reset
        // Increment counters for refreshed rows, assuming blast radius = 1
        counters[row_to_mitigate - 1]++;
        counters[row_to_mitigate + 1]++;
        mitigation.track(row_to_activate - 1);
        mitigation.track(row_to_activate + 1);
        t += PRAC_MITIGATION_TIME;
    }
    // Model checker verifies this
    assert( max(counters) <= TRH ); 
}
\end{lstlisting}

The code initializes an array of $N$ counters where $N$ is the number of rows per bank. 
At each time step $t$, the model checker non-deterministically selects a row to activate, increments its counter, and notifies the mitigation logic. 
If the defense determines a mitigation is needed and the ABO protocol permits one at the current time, a row is selected for mitigation as per its implementation logic: its counter is reset and its immediate neighbors have their counters incremented (assuming blast radius of 1). 
Time advances by tRC per activation and by ABO mitigation time when a mitigation occurs, and the loop can run until $t$ grows to tREFW.

At each step, the model asserts that no counter of the PRAC implementation exceeds \TRH{}. If the assertion fails, CBMC returns a counterexample consisting of a sequence of activations and mitigations that forms a successful attack. Starting with \TRH{} of \NBO{}+1, and by gradually incrementing the  \TRH{} by one each time, AutoPRAC can determine the safe Rowhammer threshold for a defense above which no violations occur.
However, naively running the checker over the entire tREFW is computationally infeasible. For a 32ms refresh window, the maximum number of activations per bank ($N$) can approach 550K, yielding an enormous search space of $O(R^N)$ for $R$ rows and $N$ activations. We therefore introduce two optimizations to improve tractability.

\noindent \textbf{Optimization 1: Reduce number of activations by initializing counts to \NBO$-1$.}
Prior attacks~\cite{UPRAC, QPRAC,MOAT} typically consist of a setup phase that raises counters to \NBO{}$-1$, followed by an online phase that repeatedly triggers ABO-RFMs. Since activations below \NBO{} do not affect mitigation behavior, we can safely initialize rows at \NBO{}$-1$ and verify only the attack-critical online phase, substantially reducing the number of activations simulated by the model checker.

For defenses such as MOAT~\cite{MOAT}, which additionally reset counters on refresh, attacks have an additional constraint that they must complete within a single tREFW. 
As discussed in MOAT~\cite{MOAT}, in an optimal attack, all the rows in the pool (except the last one) will trigger an ABO-RFM. 
Thus, the attack time can be calculated as the time to perform \NBO{}-1 activations per row along with the time between two consecutive ABOs (tABO) for each row,
giving $N_\text{sim} = [\text{tREFW}-8K\cdot tRFC]/ [(\text{N}_\text{BO} -1 )\cdot \text{tRC}+\text{tABO}]$. This bounds the number of rows involved in the attack and reduces the required simulation space, $N_\text{sim}$, to roughly 4000 rows for \NBO{}=128.
For defenses where counters are reset only on a mitigation, the online phase of the attack can span the entire tREFW and the number of rows to be simulated ($N_\text{sim}$) can span the entire DRAM bank. 

\noindent
\textbf{Optimization 2: Reduce number of simulated rows by storing counts in a histogram.}
We further reduce state space by exploiting symmetry between rows. Since mitigation behavior depends primarily on activation counts rather than row identities, we represent DRAM state as a histogram where \texttt{hist[i]} stores the number of rows currently at count $i$. Activating a row corresponds to moving one entry from \texttt{hist[i]} to \texttt{hist[i+1]}. This reduces the number of distinct activation choices from all rows to only the range  $[\text{N}_\text{BO} - 1, \text{T}_\text{RH}]$.

This abstraction is sound when row identity does not affect mitigation behavior. However, mitigative refreshes increment the counters of rows adjacent to the mitigated row, which is a positional effect not captured by the histogram. To model this, we maintain a second histogram, \texttt{neighbor\_bucket}, that tracks how many rows in each activation-count bucket are neighbors of previously mitigated rows. 

When a mitigation occurs, the model checker nondeterministically selects a bucket and increments its corresponding \texttt{neighbor\_bucket} entry, subject to the constraint that the number of tracked neighbors cannot exceed the total number of rows in that bucket. Similarly, when activating a row from a bucket, the model checker nondeterministically decides whether the activated row is also a tracked neighbor. This allows the model to reason about disturbance accumulation in neighboring rows without explicitly tracking row identities, while preserving the correctness of the histogram abstraction. As a result, the number of simulated variables is reduced from $O(R)$ rows to $O(2(\text{T}_\text{RH}-\text{N}_\text{BO}))$ histogram buckets.

\subsection{Modelling Counter Reset Policy on Refresh}
\label{sec:model_counter_reset}

Counter resets can introduce discrepancies between a row's actual damage and what its counter reflects. If a row's counter is reset without its neighbors being refreshed, activations that contributed to their damage become untracked. 
We model this as a separate stage of the attack. 
An attacker can first exploit counter reset to accumulate untracked damage on a victim row, and then mount a regular attack, with the total damage exceeding what the mitigation scheme accounts for.

Since rows are refreshed sequentially in groups and Rowhammer only affects spatially neighboring rows, the only groups relevant to a victim row are its own group, the group immediately before it, and the group immediately after it. 
Similarly, the only relevant time window spans the three consecutive tREFIs in which these three groups are refreshed. 
Any activations outside this window either do not affect the victim row or occur after it has already been refreshed.

Guided by this symmetry, we model three row groups and three consecutive tREFIs in which they are refreshed one by one, effectively reducing the number of simulated rows and activations. In the model, on every activation, we track the counter states and the real damage each row receives. 
The safety property we check is that, at no point does any row's real damage exceed its PRAC counter value. 

\section{Results}

\subsection{Experimental Methodology}
We implement our model using the C Bounded Model Checker (CBMC) \cite{cbmc} with Cadical \cite{cadical} as the back-end SAT solver. 
We choose CBMC as it directly operates C/C++ programs, allowing PRAC-based defenses, typically modeled in C based DRAM simulators like Ramulator2 \cite{ramulator2}, to be ported to our solver with minimal effort. 
We use a SAT back end rather than SMT since we empirically find it to be faster for our problem.



We evaluate four representative PRAC defenses: Panopticon \cite{Panopticon}, UPRAC \cite{UPRAC}, QPRAC \cite{QPRAC}, and MOAT \cite{MOAT}.\footnote{We do not evaluate Chronus~\cite{Chronus} because its open-source implementation (available at this \href{https://github.com/CMU-SAFARI/Chronus/blob/master/src/dram_controller/impl/plugin/chronus.cpp\#L264C13-L267C20}{link}, commit: 6605ffb) uses a mitigation policy that scans all row counters in DRAM to identify the maximum count, like UPRAC~\cite{UPRAC} (\href{https://github.com/CMU-SAFARI/ramulator2/blob/main/src/dram_controller/impl/plugin/prac/prac.cpp\#L266C8-L270C20}{link}), differing from the Chronus paper, which proposes using an Aggressor Tracking Table. Since the available implementation does not match the design described in the paper, we are unable to evaluate Chronus as published.}
We evaluate all four defenses with our row selection for mitigation model and only evaluate MOAT with the counter reset on mitigation model (since it is the only one that uses this policy). We use the following PRAC parameters: mitigation level 1 (mitigating one row per ABO), \NBO{} of 128, and a blast radius of 1. All experiments are run on an AMD EPYC 7713 processor a solver time limit of 2 hours and memory limit of 256GB.

For our automated attack construction process, we start by trying to verify a defense at \TRH{} of \NBO+1, and then if the solver produces a counter-example, we iteratively increase the \TRH{} by one and try again, until the solver times out.
To ensure tractability, we also limit the number of simulated rows and activations: we begin with 10 rows and set the number of activations to $N_\text{rows} \times (\text{ACT}_\text{ABO} + \text{ACT}_\text{delay})$. 
For each \TRH{}, if verification succeeds, meaning no attack is feasible given the pool size, we increase the number of rows by 5 and repeat, until either a violation is found or the verification times out.

\subsection{Experimental Results}



\begin{figure}[htbp]
\centering
\includegraphics[width=\columnwidth]{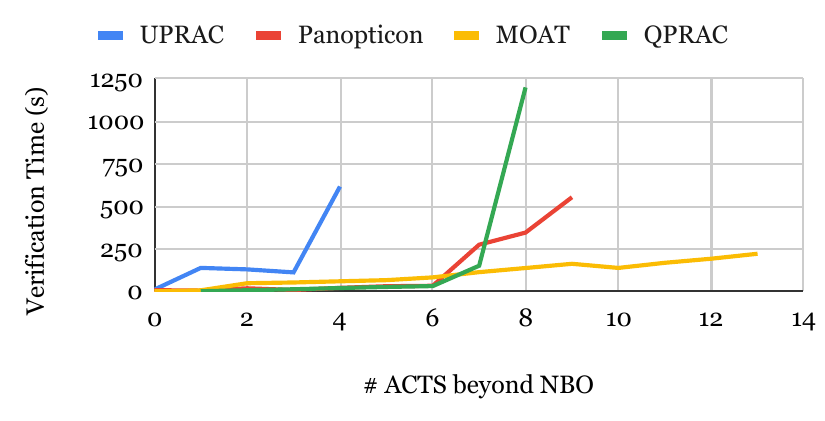}
\caption{Verification time of the model for the mitigation policy (queue design) vs. number of activations above \NBO{}.}
\label{fig:time_to_verify}
\end{figure}

\textbf{Performance Analysis.} 
Since we incrementally verify the behavior of defenses at \TRH{} values starting from \NBO{} $+1$, the verification complexity scales with the number of activations simulated above \NBO{}. 
\Cref{fig:time_to_verify} shows verification time for the model of the mitigation policy (queue design) across defenses as number of \ACTs{} above \NBO{} increases. 
For UPRAC, Panopticon and QPRAC, there is a sharp increase in verification time at the point where the number of simulated rows grows beyond the initial pool size, to verify increasing activation counts.
For UPRAC, the verification time grows much faster, as there are no queues to track mitigation, making the solver slower since it needs to traverse a larger space to identify the highest activated row at any time.
In comparison, for QPRAC and Panopticon, the queue limits the search space for the solver to select the row to be mitigated, reducing the runtime relative to UPRAC. 
For MOAT, verification time grows more gradually since it only tracks a single row for mitigation, but the solver times out beyond 13 activations. 
Across all defenses, despite the 2 hour limit, the verification runs tend to either complete within 20 minutes or time out, with little in between. 
This reflects the exponential growth in search space with SAT solvers. We also observe that increasing the time limit to 24 or 48 hours does not meaningfully improve the number of ACTs above \NBO{} for a counter example.
In comparison, the model for the counter-reset policy completes within minutes.

\begin{figure}[htbp]
\centering
\includegraphics[width=\columnwidth]{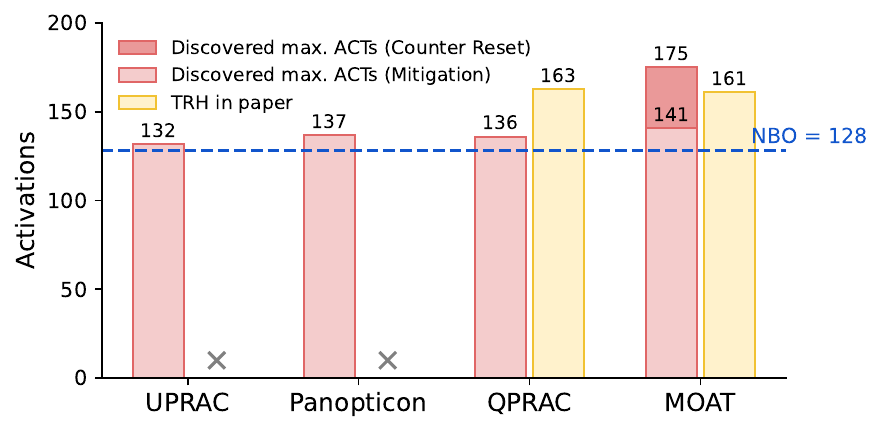}
\caption{Maximum activations to a row in attacks discovered by AutoPRAC at \NBO{} of 128, compared to \TRH{} in prior papers (we could not find the \TRH{} at NBO of 128 in Panopticon and UPRAC paper). AutoPRAC discovers a stronger attack on MOAT than previously known by exploiting counter resets.}
\label{fig:max_verified}
\end{figure}

\paragraph{Maximum Activation Count in Discovered Attacks.} \Cref{fig:max_verified} shows the maximum activation counts for an attacked row, automatically discovered by AutoPRAC (for \NBO{} of 128) across defenses, compared to the \TRH{} claimed in the respective papers. For the mitigation selection model, AutoPRAC's verification can only find counter examples till 4 to 13 \ACT{}s above \NBO{}, beyond which it times out. Hence, it can automatically find attacks with only a maximum of 132 to 141 \ACTs{} to a row, for UPRAC, Panopticon, and QPRAC, which are below human-developed attacks in these respective papers that reach till a \TRH of 163. 
However, for MOAT, AutoPRAC finds an attack that reaches 175 \ACT{}s, higher than the \TRH{} of 161 \ACT{}s claimed by MOAT~\cite{MOAT}. Thus, AutoPRAC finds new attacks that defeat MOAT's security guarantees that were previously based on human-developed attacks.  

Of the 47 \ACT{}s above \NBO{} that AutoPRAC produces in its counter-example, 13 \ACT{}s are based on its mitigation policy model, and 34 \ACT{}s are based on its counter-reset policy model.
The counter-reset vulnerability we discover is orthogonal to the optimal human-developed attack proposed by MOAT, and the two can be combined. Together, they induce 195 ACTs on a row prior to mitigation (34 activations above the TRH of 161 claimed by MOAT \cite{MOAT}) making any TRH of 195 or below insecure.
We describe the counter-reset vulnerability discovered by AutoPRAC in more detail next.

\subsection{MOAT's Bug in Counter Resets on Refresh}
\label{sec:case_study}

\begin{figure}[htbp]
\centering
\includegraphics[width=\columnwidth]{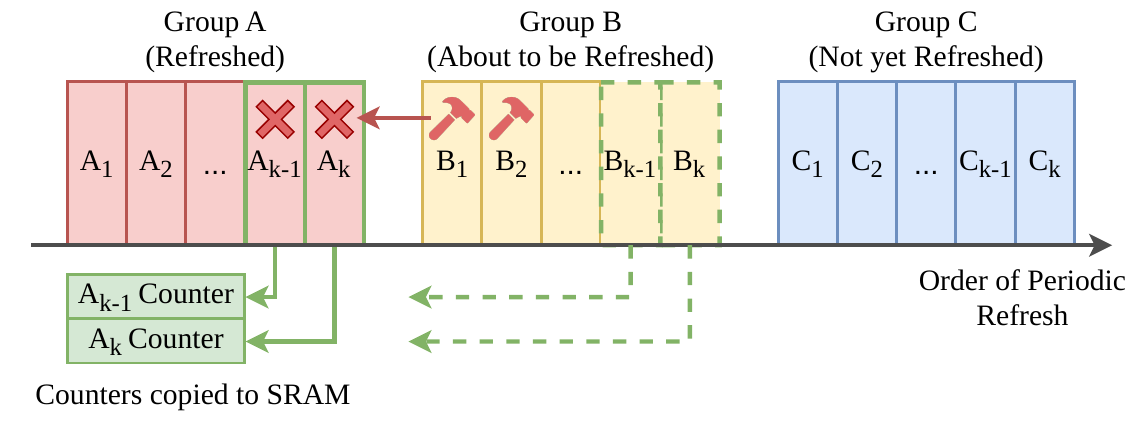}
\caption{MOAT's counter reset policy on refresh~\cite{MOAT} and AutoPRAC's new attack hammering to be refreshed rows ($B_1, B_2$). }
\label{fig:moat_bug}
\end{figure}

MOAT's \cite{MOAT} counter reset policy is shown in \Cref{fig:moat_bug}. 
It divides rows in a bank into consecutive groups of $k$ rows, sequentially refreshing one group every tREFI. 
Assuming the blast radius of hammering is 2, they noted that only the last two rows of a recently refreshed group (e.g., $A_{k-1}$ and $A_k$ in group A) can pose threat to neighbors in the next group B, and therefore their counter values need to be stored separately in SRAM until the next group is refreshed; All other rows in the recently refreshed group have their counter values reset to 0.
This ensures that no tracking information is lost in the interim between refresh of groups A and B.

AutoPRAC provided a new attack on MOAT automatically, as a counter-example,
which allows 34 activations on MOAT beyond its \TRH{} while evading mitigation.
This attack exploited the fact that while MOAT protects victim rows in B from being hammered by the last two rows in Group A in the gap between refresh to Group A and Group B, it does not protect victim rows in A from being hammered by rows of B. 
In this attack we discovered, after group A's rows are
refreshed, the attacker activates the first two rows of B alternately
($B_1$,$B_2$,$B_1$,$B_2$,$\cdots$) until tREFI ends or their counts reach
$\text{N}_\text{BO}-1$, for a total of $k$ activations. The refresh of group B rows resets the
counters of $B_1$ and $B_2$ to 0, but row $A_k$, in the blast radius of $B_1$
and $B_2$ has disturbance equivalent to $k$ activations which now goes untracked after the reset of rows $B_1$ and $B_2$ after the refresh of group $B$. Starting from zeroed counters and assuming
67 activations per tREFI in DDR5~\cite{MOAT}, the attacker accumulates up to
$\min(34, \text{N}_\text{BO} - 1)$ untracked activations
on $B_1$ and $B_2$ each; this surpasses the best known attack on MOAT by 34 activations and
breaks MOAT's security.

This bug is symmetric to a case that MOAT's designers already handled at
the end of a refresh group. A simple fix to the bug is to track the activation count for rows $B_1$ and $B_2$ in SRAM, between the time of refresh of rows in Group A and Group B. After the refresh of Group B, the counts for these two rows in DRAM is reset to the SRAM counts, to ensure no information loss due to staggered counter resets on refresh. 

More broadly, the case study shows that even when designers are aware of a class of
vulnerability, manually reasoning about all attack variants is
challenging. AutoPRAC surfaced this bug automatically, demonstrating
its value as a automated attack discovery tool for PRAC designs.

\section{Limitations and Future Work}

\noindent
\textbf{Scalability.}
AutoPRAC only produces attacks around 2000 activations in length, whereas real Rowhammer attacks can involve hundreds of thousands of activations within a tREFW. The main bottleneck is solver scalability: as the number of simulated rows and activations increases, the search space grows exponentially, eventually causing solver timeouts.
Future works could improve scalability in several directions.

First, efficient problem encoding could reduce solver complexity and improve its performance. For example, our abstraction of row count histograms reduced the number of simulated rows by abstracting row state into buckets. Future works may explore more efficient abstractions may for the attack sequence, currently represented as a flat list of activations, that could enable more effective search-space pruning by the solver.

Second, the choice of model checker and modeling language may affect performance. While CBMC's C/C++ interface simplifies portability, it cannot exploit the domain-specific optimizations available in dedicated modeling languages. Tools such as Spin~\cite{spin} provide native support for data structures like queues, richer type systems, and solver optimizations that may make reasoning about PRAC designs more efficient.

Third, because AutoPRAC is used for attack generation, performance depends not only on solver speed but also on how quickly a counterexample is found. Reformulating constraints or reordering verification goals to steer the search toward promising attack patterns could substantially reduce runtime without modifying the underlying solver.

Finally, AutoPRAC could be combined with complementary techniques such as fuzzing or human-guided attack construction to accelerate counterexample discovery. Together, these improvements could help AutoPRAC match or surpass the effectiveness of human-crafted attacks and move closer to full verification of PRAC defenses.

\smallskip
\noindent
\textbf{Model Completeness.}
Verification guarantees are inherently limited by the completeness of the model. Our model assumes a fixed Rowhammer blast radius, whereas prior work such as SALT~\cite{SALT} shows that disturbance can decay gradually across distant rows, enabling Ripple Attacks in PRAC defenses. Modeling such effects in solvers is challenging because it requires reasoning about fractional and exponentially decaying disturbance, which SAT/SMT solvers handle inefficiently.
Our model also focuses on the core PRAC protocol and does not capture optimizations proposed by prior work, such as proactive mitigation on REF commands~\cite{QPRAC}, dual thresholds~\cite{MOAT,QPRAC} for energy efficiency, or victim activation counting~\cite{PVAC}. 
Future works can extend AutoPRAC to include such modeling details.
Beyond Rowhammer, DRAM is also susceptible to other read disturbance phenomena. RowPress~\cite{RowPress} exploits long periods for which a row is kept open rather than rapid activations, and ColumnDisturb~\cite{ColumnDisturb} affects rows in neighboring subarrays rather than just neighboring rows. 
Future works could extend AutoPRAC to verify defenses against these broader classes of DRAM disturbance attacks.
Finally, future work could apply formal verification directly at the RTL level, reducing the gap between algorithmic design and hardware implementation and enabling detection of RTL-level bugs.

\section{Conclusion}
We present AutoPRAC, the first automated framework for attack discovery in PRAC-based Rowhammer defenses using bounded model checking. AutoPRAC models key security-sensitive aspects of PRAC designs, including mitigation policies and counter-reset behavior, and automatically synthesizes attack traces when security properties are violated.
While solver scalability currently prevents AutoPRAC from matching the longest human-crafted attacks in three of the four evaluated defenses,
AutoPRAC uncovers a previously unreported flaw in MOAT's counter-reset policy that allows 34 activations beyond the claimed Rowhammer threshold to go untracked. 
Our results thus demonstrate the value of AutoPRAC as an automated attack-discovery tool for PRAC defenses, and we hope it serves as a foundation for more scalable attack discovery and comprehensive verification of future Rowhammer mitigations.

\bibliographystyle{IEEEtran}
\bibliography{refs}
\balance
\vspace{12pt}
\end{document}